\documentclass{aa} 

\usepackage{graphicx}
\usepackage{txfonts}
\begin{document}

 \title{Comparing the spatial and kinematic distribution of gas and young stars around the shell-like structure in the CMa OB1 association}

 \subtitle{}

 \author{J. Gregorio-Hetem\inst{1}
 \and
 B. Lefloch\inst{2}
 \and
 A. Hetem\inst{3}
 \and
 T. Montmerle\inst{4}
 \and
 B. Fernandes\inst{1}
 \and
 E. F. Mendoza\inst{1,5}
 \and
 M. De Simone\inst{2,6}
 }

 \institute{
 Universidade de S\~ao Paulo, IAG, Rua do Mat\~ao 1226, 05508-090 S\~ao Paulo, Brazil
 \email{gregorio-hetem@usp.br}
 \and
 Univ. Grenoble Alpes, CNRS, IPAG, 38000 Grenoble, France 
 \and
 UFABC Federal University of ABC, Av. dos Estados, 5001, 09210-580 Santo Andr\'e, SP, Brazil 
 \and
 Institut d'Astrophysique de Paris, 75014, Paris, France
 \and
 Observat\'orio do Valongo, UFRJ, 20080-090 Rio de Janeiro RJ, Brazil
 \and
 INAF, Osservatorio Astrofisico di Arcetri, Largo E. Fermi 5, 50125 Firenze, Italy
 }

 \date{Received --; accepted --}

 
 \abstract
 {The relationship between young stellar clusters and respective parental molecular clouds is still an open issue: for instance, are the similarities between substructures of clouds and clusters just a coincidence? Or would they be the indication of a physical relationship? In order to address these issues, we have studied the CMa OB1/R1 region that shows evidence for a complex star formation history. }
 {We obtained molecular clouds mapping with the IRAM-30 metre telescope to reveal the physical conditions of an unexplored side of the CMa region aiming to compare the morphology of the clouds with the distribution of the young stellar objects (YSOs). We also study the clouds kinematics searching for gradients and jet signatures that could trace different star formation scenarios.}
 {The YSOs were selected on the basis of astrometric data from {\it Gaia} EDR3 that characterise the moving groups. The distance of 1099$_{-24}^{+25}$ pc was obtained for the sample, based on the mean error-weighted parallax.
 Optical and near-infrared photometry is used to verify the evolutionary status and circumstellar characteristics of the YSOs.}
 {Among the selected candidates we found 40 members associated with the cloud: 1 Class I, 11 Class II, and 28 Class III objects. Comparing the spatial distribution of the stellar population with the cores revealed by the 
 $^{13}$CO map, we verify that peaks of emission coincide with the position of YSOs confirming the association of these objects to their dense natal gas.}
 {Our observations support the large-scale scenario of the CMa shell-like structure formed as a relic of successive supernova events.}

\titlerunning{The spatial and kinematic distribution of gas and young stars in the CMa OB1 association}
\authorrunning{Gregorio-Hetem et al.}
\maketitle
%

\section{Introduction}

 OB associations are ideal sites to test our understanding of star formation and how this process is influenced by the feedback from massive stars. The interplay between supernova (SN) events and the star-forming molecular cloud is of key relevance to the star formation process as shown by several examples. In particular, it has been proposed that SN are able to affect star formation negatively by suppressing the formation of new stars in their surroundings, and positively by triggering it \citep[see review by][]{Hensler11}. In the latter case, the expansion of SN remnant (SNR) shells can sweep up the surrounding gas up to the point of triggering sequential star formation \citep[according to the ``collect and collapse" model of][]{EL77}. 

The extended HII region Sh~2-235, for instance, is an active star-forming region, where star formation triggered by a SNR seems to have occurred in two nebulae: S235A and S235B \citep{Kirsanova14}. The young stars associated with both nebulae are $\sim$0.3~Myr old, coinciding with the age of the SNR proposed by \citet{Kang12}. Another interesting example is W28 SNR \citep{Lefloch08,Vaupre14}, whose interaction with molecular clouds could have triggered the formation of nearby protostellar clusters in the Trifid nebula. In the case of the SNR IC443, however, although many young stellar objects (YSOs) are found surrounding the SNR shell, the SNR proved to be too young as compared to the age of YSOs and could not have triggered their formation \citep{Xu11}. The recent numerical simulations by \citet{Dale15} could easily reproduce triggering of star formation. However, when comparing with observations the authors show that triggered star formation is much harder to infer, since they could not discriminate triggered from non-triggered objects.

Star formation is, in any case, often found nearby SNRs, for instance there are several SNRs associated with YSOs in the Large Magellanic Cloud \citep[e.g.][]{Desai10}, and supernovae have been long suggested to be exciting the star formation in their surroundings. The arc-shaped Sh~2-296 nebula is one of these cases, which is suspected to be an old SNR that could have triggered the formation of new stars in CMa OB1 \citep{Herbst77}.

The Canis Major OB1/R1 (henceforth CMa, for simplicity) is a nearby (d $\sim$ 1 kpc) OB Association with a complex star formation history. Our previous studies showed that the region contains young objects originated from different star-forming events \citep{GH09,SS18}.

\begin{figure*}
\centering
\includegraphics[width=0.8\textwidth,angle=0]{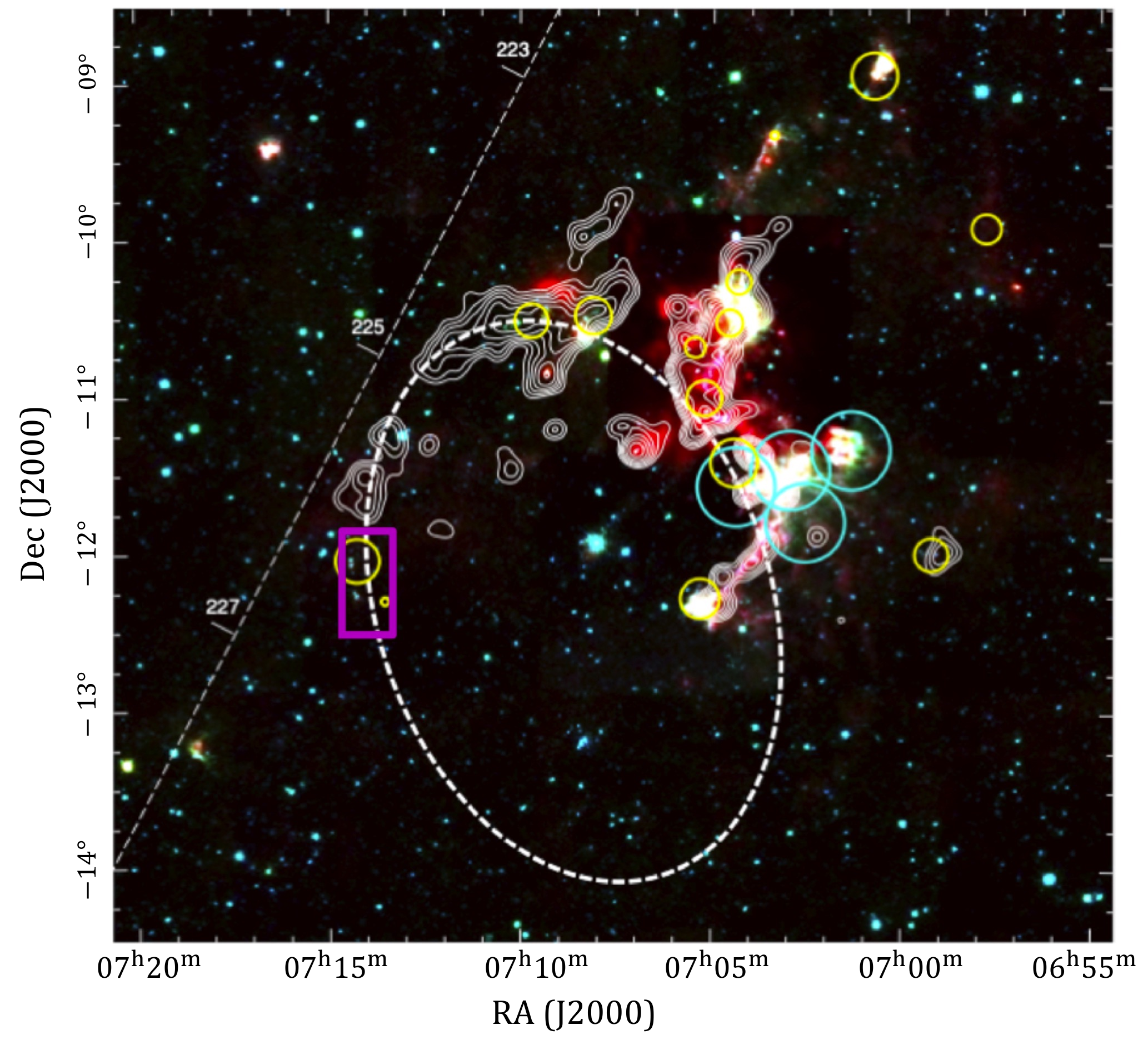} 
\caption{ Overview of the CMa R1 region. The background is a RGB composite WISE image (Red: 22.2 $\mu$m, Green: 12.1 $\mu$m, Blue: 3.4 $\mu$m). A dashed straight line gives longitudes in the Galactic Plane, and the dashed ellipse shows the area of the CMa shell suggested by \citet{F19}. White contours trace the molecular gas ($^{13}$CO). X-ray observations with {\it XMM-Newton} \citep{SS18} are indicated by cyan circles, and the groups of YSOs studied by \citet{Fischer16} are shown by yellow circles. The area studied here is marked by the magenta box. \citep[Figure adapted from][]{F19}.}
\label{fig:overall}
\end{figure*}


A promising hypothesis is presented by \citet{F19} showing new evidence that Sh~2-296, the most prominent nebula in the CMa Association, is part of a large (diameter $\sim$ 60 pc) shell-like structure. Figure \ref{fig:overall} shows this structure, called the {\it CMa shell}, that likely results from successive SN explosions from 6 to 1 Myr ago as inferred from the past trajectories of three runaway stars in the region, derived from {\it Gaia} proper motions. They also found evidence that the {\it CMa shell} is related to a larger ($\sim$ 140 pc in size) shell structure, visible in H$\alpha$.
However, the older population of low-mass stars (> 10 Myr), which are confirmed CMa members, cannot be explained by these recent SN explosions suggesting that they may be causally related to the existence of the H$\alpha$ super-shell.

\citet{F19} suggest that the present-day configuration of the star-forming gas may have been shaped by a few successive SN explosions that occurred several Myr in the past.
Despite the evidence of SN events shaping the CMa shell, \citet{F19} argue that they probably played a minor role in triggering star formation in these clouds.

CMa is, therefore, an ideal laboratory for probing how the feedback from SNRs interacting with molecular clouds can affect their environment and subsequent star formation and evolution in OB associations. 

\section{Molecular clouds in CMa}
\label{sec:clouds}

A $^{13}$CO (J = 1-0) survey of \citet{Kim04}, using the Nagoya-4m telescope with a beam width of 2\rlap{.}$^{\prime}$7, identified 13 
molecular clouds in the area of the CMa Association, distributed in three main structures around the clouds 
No. 3 (l$\sim$224$^{\rm o}$, b$\sim-$2$^{\rm o}$), No. 4 (l$\sim$224\rlap{.}$^{\rm o}$5, b $\sim-$1$^{\rm o}$), 
and No. 12 (l $\sim$226$^{\rm o}$, b$\sim-$0\rlap{.}$^{\rm o}$5).

Partial maps of these molecular clouds were recently obtained with the 1.85m mm-submm Telescope 
 installed at the Nobeyama Radio Observatory\footnote{As courtesy of the Osaka University group, a $^{13}$CO map of the CMa region was obtained by T. Onishi and K. Tokuda (private communication).}.
 According to \citet{Onishi13}, the 1.85m telescope is dedicated to a large-scale survey aiming to reveal the physical properties of molecular clouds in the Milky Way Galaxy. In the 1.3mm band, observations of the rotational transition $J$= 2--1 of $^{12}$CO, $^{13}$CO and C$^{18}$O were obtained with a beam size of 2\rlap{.}$^{\prime}$7.

\begin{figure*}
\centering
\includegraphics[width=0.9\textwidth,angle=0]{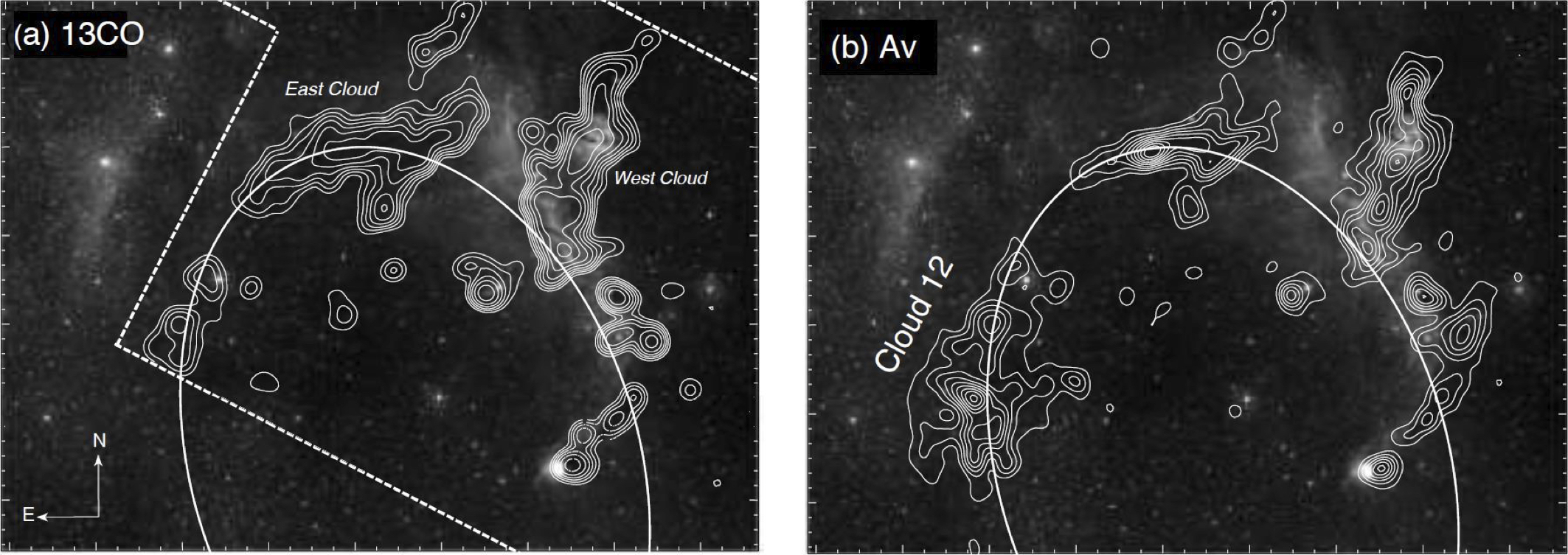} 
\caption{DSS2-Red image of the Sh~2-296 nebula. The ellipse marks the proposed shape of the CMa shell and the contours in each panel trace: (a) the $^{13}$CO emission (courtesy of the Osaka group, see \citet{Onishi13}); (b) extinction ($A_V$) from the 2MASS maps of \citet{Dobashi11}. \citep[Figure adapted from][]{F19}.}
\label{fig:avmap}
\end{figure*}

The $^{13}$CO map shows a chain of molecular clouds that extends North-East of the Sh~2-296 nebula. 
 In Fig.~\ref{fig:avmap}a, the main structure associated with Sh~2-296 is called ``West cloud", which coincides with ``Cloud 3" of the list from \citet{Kim04}. This is the largest cloud of the region with mass $\sim$ 16000 M$_{\odot}$ and an area of 358 pc$^2$. 
To the North-East, the second main structure, called ``East cloud", is related to ``Cloud 4" 
(total mass $\sim$ 12000 M$_{\odot}$, and area $\sim$ 301 pc$^2$).

The clouds are also traced by the dust distribution revealed by the extinction map of A$_V$ (Fig. \ref{fig:avmap}b) from \citet{Dobashi11}. We can see that the dust emission matches closely the $^{13}$CO distribution, following the approximate shape of the CMa shell. Comparing Figs. \ref{fig:avmap}a and \ref{fig:avmap}b, it can be noted the lack of information about $^{13}$CO emission for ``Cloud 12", which is found in the area not covered in the survey obtained by the Osaka group (see the dashed rectangle in Fig. \ref{fig:avmap}a).
This cloud is the third largest gas reservoir in the CMa association, with an estimated mass of 7500 M$_{\odot}$ and an area of 315 pc$^2$ \citep{Kim04}. Besides its important amount of gas, this cloud is located in the border of the {\it CMa Shell}, in the opposite side of Sh~2-296 nebula. 

This work is dedicated to investigate this complementary area of gas distribution, which can bring an important contribution to understand the star formation scenario in CMa.
In the general context, we aim to explore the global dynamics that allows us to search for signatures of the large-scale SNR driven shock and its interaction with the molecular gas condensation. 
In Sect. \ref{sec:obs}, we summarise the surveys used in the analysis and the molecular gas observations. The identification and analysis of young stars associated with the cloud are presented in Sect. \ref{sec:stars}. Finally, in Sect. \ref{sec:compara} we discuss the results from the molecular clouds mapping in comparison with the characteristics of the associated stellar population. The conclusions are summarized in Sect. \ref{sec:conclude}.

\section{Observational data} 
\label{sec:obs}
\subsection{Dust continuum surveys}
\label{sec:dust}

Besides the visual extinction map that was compared with the molecular clouds distribution in Sect. \ref{sec:clouds}, we analyze here the dust distribution traced by the infrared emission. We are particularly interested in searching for condensations and filamentary structures aiming to verify whether they are harbouring pre- and protostellar cores, as suggested by \citet{Elia13}.

As part of the {\it Herschel Infrared Galactic plane survey} \citep[Hi-GAL][]{Molinari10}, \citet{Elia13} conducted a study 
 of star formation in the third Galactic quadrant, which includes the CMa region. In such study,
{\it Herschel} PACS and SPIRE\footnote{Instruments on the Herschel Space Observatory \citep{Pilbratt10}: PACS (Photodetector Array Camera \& Spectrometer) \citep{Griffin10} covers the 70 and 160 $\mu$m bands, while SPIRE (Spectral and Photometric Imaging REceiver \citep{Poglitsch10} operates at 250, 350, and 500 $\mu$m.}
 photometric observations were combined with NANTEN CO $J$=1--0 observations of cores and clumps, revealing that most of the protostars are in the early accretion phase, while star formation is still underway in cores distributed along filaments. 

\begin{figure*}
 \centering
\includegraphics[width=0.48\textwidth,angle=0]{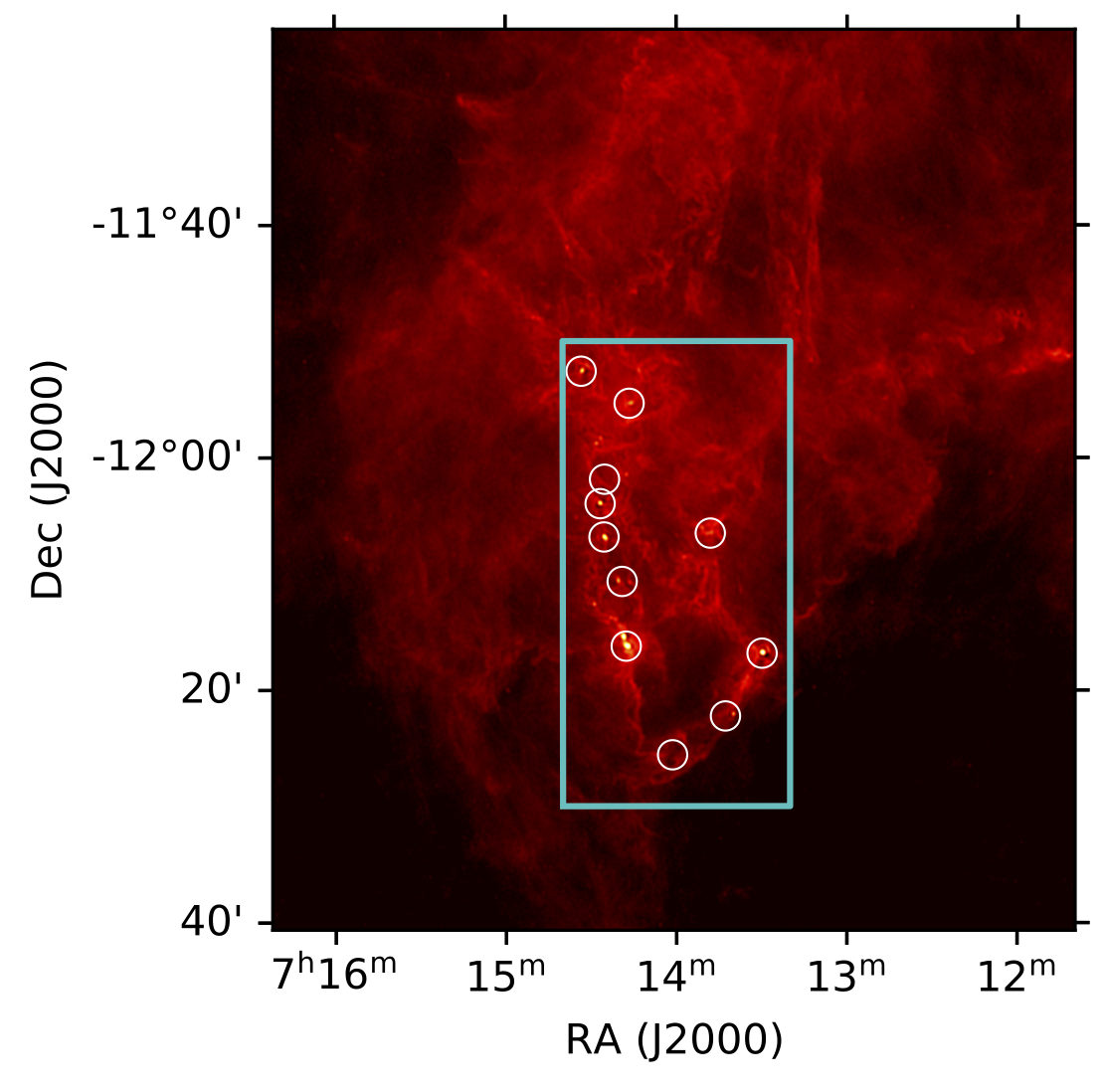} 
\includegraphics[width=0.48\textwidth,angle=0]{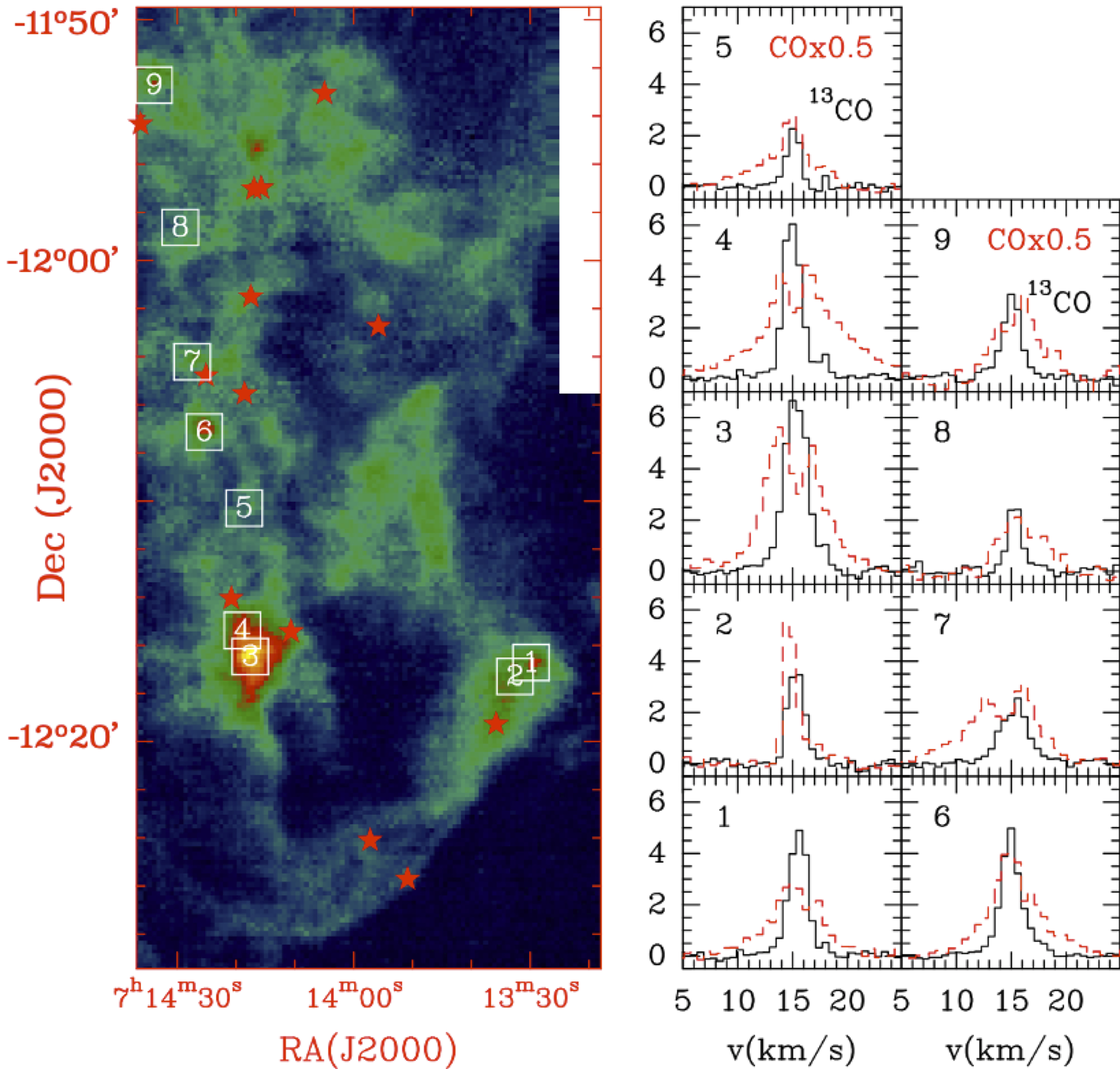} 
\caption{{\it Left}: The region covered by the IRAM observations (cyan box) overlaid on the {\it Herschel} SPIRE 250$\mu$m map that illustrates the filamentary structure of the cloud and the distribution of bright IRAS sources (white circles).
{\it Middle:} $^{13}$CO map revealing dense cores roughly related to the IRAS sources. The location of the WISE 
sources (Class I) are marked by red stars. The location of the 9 dense cores listed in Table \ref{tab:coeurs} is drawn with white squares. 
{\it Right:}
The core emission ($T_{MB}$) observed in the $^{13}$CO (black) and $^{12}$CO (dashed red) $J$=1--0 transitions. A scaling factor of 0.5 has been applied to the $^{12}$CO line intensity. }
\label{fig:spire}
\end{figure*}

The filamentary structure of ``Cloud 12" studied here can be clearly seen in the {\it Herschel} SPIRE map obtained at 250 $\mu$m, which is shown in Fig. \ref{fig:spire} (left panel). The position of bright infrared sources (flux density $>$ 3 Jy at 100 $\mu$m band) from the {\it IRAS Catalogue of Point Sources} \citep{iras} is also plotted in this figure, showing a good correlation with the bright filaments. 
 
 We also performed the characterisation of the IR sources associated with the clouds by searching for candidates in the
 {\it AllWISE} catalogue \citep{Cutri13}. The WISE photometry \citep{wise} at bands W1 (3.4 $\mu$m), W2 (4.6 $\mu$m), and W3 (12 $\mu$m)
 are useful to distinguish different classes of YSOs as a function of their IR excess.
 
 As shown in Fig. \ref{fig:overall}, the CMa clouds contain several groups of YSO candidates identified by \citet{Fischer16}
based on WISE colors. The candidates were selected by adopting the criteria proposed 
by \citet{K14} to identify Class I and Class II objects. Details on this method, which is also applied by us, is presented in Sect. \ref{sec:wise}. The distribution of the groups of YSOs found by \citet{Fischer16} closely follows the border of the CMa Shell and coincides with the gas distribution.

\begin{table*}
\caption{Observational properties of a sample of molecular cores. }
\begin{tabular}{crrccccccc} 
\hline
Core & $\Delta\alpha$& $\Delta\delta$ & RA (J2000) & Dec (J2000) & T$_{kin}$ & N($^{13}$CO) & N(H$_2$) & outflow & comment \\
 & (arcsec) & (arcsec) & ( $^{\rm h}$~$^{\rm m}$~$^{\rm s}$ ) &( $^{\circ}$~ $^{\prime}$~ $^{\prime \prime}$ ) & (K) & ($10^{16}$cm$^{-2}$) & ($10^{22}$cm$^{-2}$) & & \\ 
\hline
1 & $-$325 & 165 &07~13~30 &$-$12~16~43 & 8 & 3.3 & 2.1 & Y & \\ 
2 & $-$285 & 130 &07~13~33 &$-$12~17~18& 8 & 3.3 & 2.1 & N & \\ 
3 & 375 & 180 &07~14~17 & $-$12~16~28& 13 &5.0 & 3.3 & Y & Broad wings \\ 
4 & 395 & 245 &07~14~18 &$-$12~15~23 & 13 &3.5 & 2.3 & Y & Broad wings \\ 
5 & 390 & 550 &07~14~18 &$-$12~10~18 & 10 & 3.8 & 2.5 & Y & Broad wings \\ 
6 & 490 & 740 &07~14~25 &$-$12~07~08 & 12 & 1.2 & 0.8 & Y & Small wings \\ 
7 & 520 & 915 &07~14~27 &$-$12~04~13 & 12 & 0.7 & 0.5 & Y & Blueshifted wing \\ 
8 & 550 &1250 &07~14~29 &$-$11~58~38 & 7 & 0.7 & 0.5 & N & \\ 
9 & 615 &1605 &07~14~33 &$-$11~52~43 & 9.5 &1.2 & 0.8 & N & \\ 
\hline
\end{tabular}
Note: The physical parameters: kinetic temperature (T$_{kin}$), and column density (N($^{13}$CO)) were obtained from 
averaging the signal over a region of $20^{\prime\prime} \times 20^{\prime\prime}$. A value T$_{kin}$ = 10K was adopted for Core 5 due to the lack of detection of C$^{18}$O.
\label{tab:coeurs}
\end{table*}


\subsection{Molecular gas}

In order to characterise the properties of the cores and filaments associated with the star formation activity, as revealed in the {\it AllWISE} and {\it Herschel} surveys, we have used the IRAM-30m telescope (Sierra Nevada, Spain) to map the emission of the ground state rotational transition $J$=1--0 of CO and its rare $^{13}$CO and C$^{18}$O isotopologues in Cloud 12. Observations were carried out during three observing runs in October 2018 (project 043-18), March 2019 (project 120-19) and October 2019 (project 034-20), using the EMIR receiver at 3mm in its 2SB mode connected to the 
FTS spectrometer in its 192 kHz resolution mode. Observations were carried out using the ``On-The-Fly" mode. We chose a reference position, which we checked to be free of emission in the CO $J$=1--0 line. 
We mapped a total area of 20$^{\prime}$ $\times$ 40$^{\prime}$ in the rotational transitions $J$=1--0 of CO, $^{13}$CO and C$^{18}$O. 
Figure \ref{fig:spire} (left panel) shows the mapped area superimposed upon an image of the continuum emission from Cloud 12 obtained by {\it Herschel} SPIRE at 250 $\mu$m (see Sect. \ref{sec:dust}). 

The weather conditions were good and rather stable during the observing sessions. Atmospheric calibrations were performed every 12 to 15 min and showed the weather to be stable. Pointing was monitored every hour on a nearby quasar and corrections were always found lower than 3$^{\prime\prime}$. Special attention was paid to the line calibration and we obtained a very good agreement between the different runs. 
The calibration uncertainty is about 10\% in the 3mm band. 

The data reduction was performed using the GILDAS software developed at IRAM\footnote{http://www.iram.fr/IRAMFR/GILDAS/}. 
The line intensities are expressed in units of antenna temperature corrected for atmospheric attenuation and rearward losses ($T_A^*$). For the subsequent radiative transfer analysis of the pre- and protostellar core emission, line fluxes were expressed in units of main beam temperature ($T_{MB}$). The main beam efficiency and the half power beam width (HPBW) were taken from the IRAM 
webpage\footnote{http://publicwiki.iram.es/Iram30mEfficiencies}.

The right panel of Fig.~\ref{fig:spire} illustrates the pre- and protostellar condensations, 
that were detected thanks to the IRAM-30m angular resolution. Table \ref{tab:coeurs} gives the results for a sample of molecular cores shown in Fig. \ref{fig:spire} (middle panel).


\section{Stellar population}
\label{sec:stars}

Two groups of YSOs \citep{Fischer16} are projected unto the eastern border of the {\it CMa shell}, coinciding with the location of the ``Cloud 12" identified by \citet{Kim04}. Our preliminary analysis on the distribution of YSO candidates (selected from the \textit{AllWISE} catalogue) is well correlated with the dust distribution revealed by the A$_V$ map in a 800 arcmin$^2$ area within the cloud (Fig. \ref{fig:avmap}). 

Here, we extracted optical and infrared data (public catalogues) for stars found in the direction of the fields observed with IRAM-30m, searching for candidates that probably 
are members associated to the cloud (Sect. \ref{sec:gaia}). The selected members are then characterized, based on infrared colors that allow us to identify Classes I and II 
objects (Sect. \ref{sec:wise}). The confirmation of the pre-main sequence nature of the candidates is obtained from color-magnitude diagram using {\it Gaia} EDR3 photometry 
(Sect. \ref{sec:age}).


\subsection{Selection of members}
\label{sec:gaia}

Aiming to exclude the presence of field-stars in the sample, as well as confirming the membership of the objects associated with the cloud, we performed a selection of kinematic members by using the techniques described by \citet{HGH19}. 
Following the formalism presented by \citet{Dias14}, the adopted statistical methods use likelihood model and cross entropy technique to estimate the probability of a candidate to be (or not to be) considered a cluster member.
A vector of parameters consisting of astrometric and kinematic data given by the observed proper motion is used to calculate the probability density function for a candidate and for the background of field-stars. 
The membership probability basically results from the fitting of a Gaussian in 5D phase space (three positions and two components of proper motion) by comparison with the Gaussian background. 
Since the cross entropy is sensitive to the initial parameters, a genetic algorithm code is adopted for parameters optimisation \citep{HGH19}.

The first subset of candidates was obtained by querying astrometric and kinematic data from the {\it Gaia} EDR3 catalogue \citep{G2,G3}. The search was performed in an area
slightly larger than the fields observed with IRAM-30m. We also delimited the query in ranges of parallax and proper motion compatible with results previously found for CMa \citep[e.g.][]{SS20}. According to the {\it Gaia} technical recommendations\footnote{http://www.rssd.esa.int/doc\_fetch.php?id=3757412},
we applied the RUWE{\footnote{Re-Normalized unit weight error (see details in the technical
note GAIA-C3-TN-LU-LL-124-01)}} $< 1.4$ and $\varpi / \sigma_{\varpi} > 3$ selection filters, in order to avoid low quality of the astrometric solution.

\begin{figure}
\begin{center}
\includegraphics[width=0.21\textwidth]{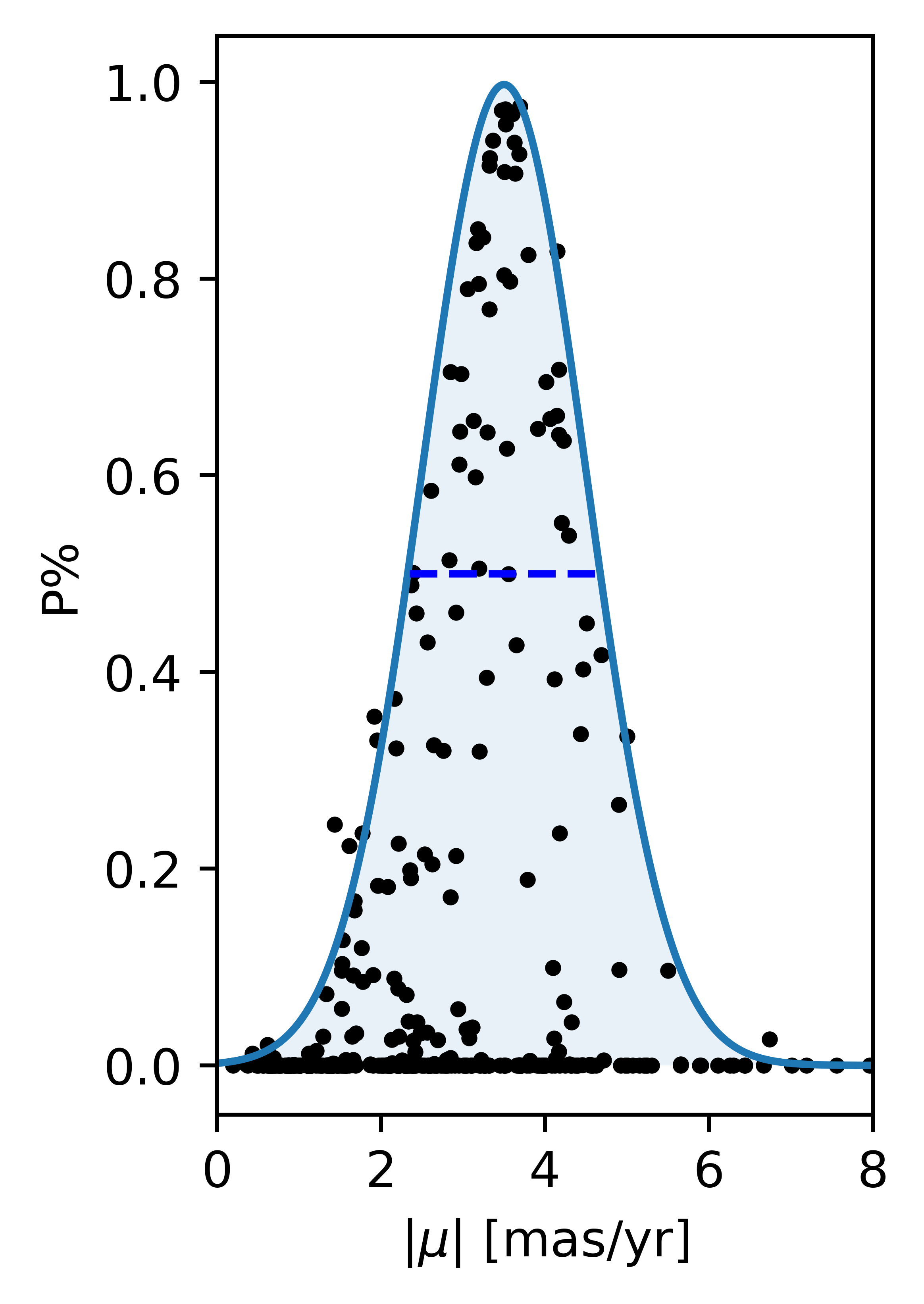} 
\includegraphics[width=0.27\textwidth]{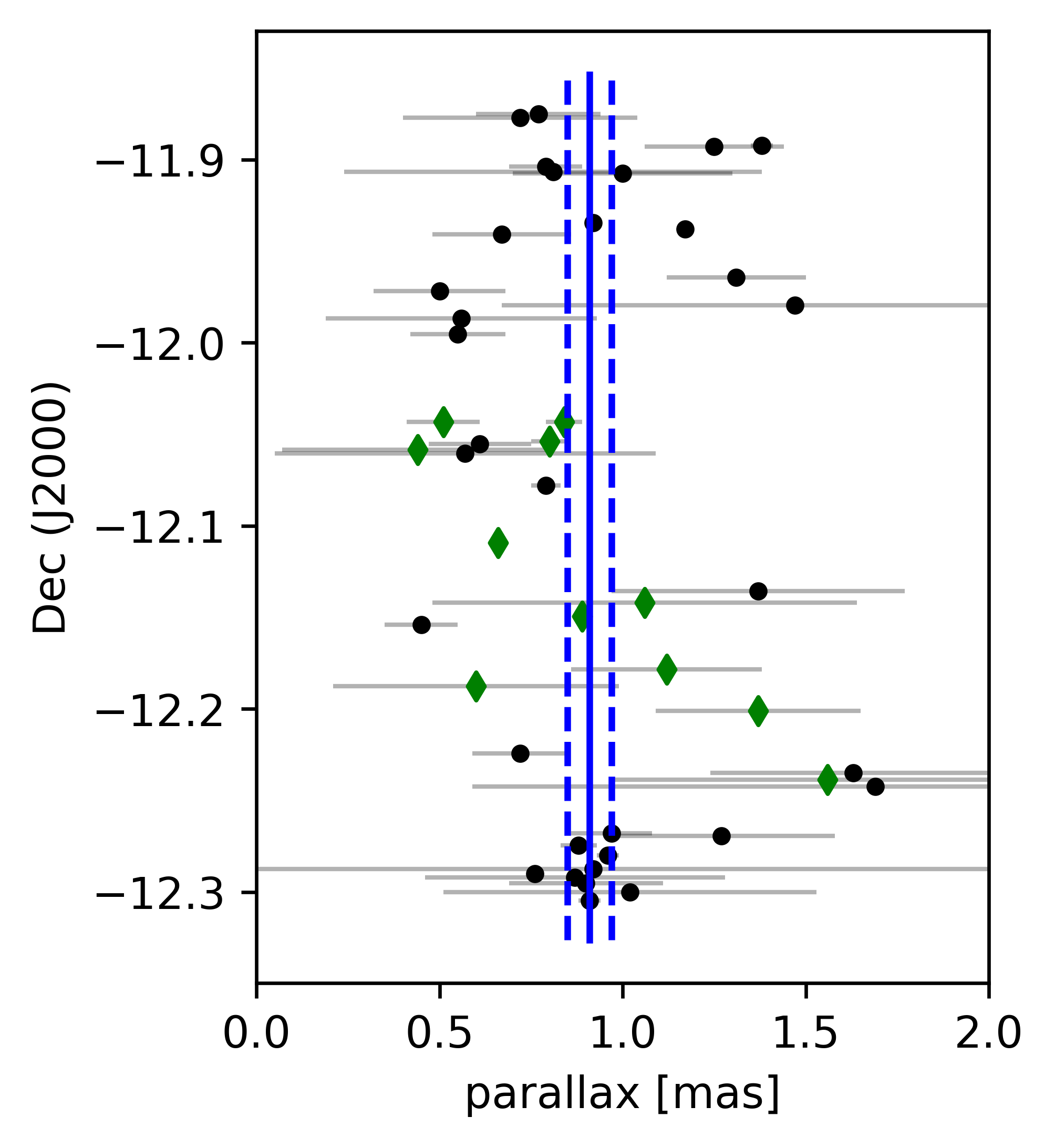} 
\end{center}
\caption{{\it Left:} Distribution of membership probabilities as a function of modulus of proper motion. A dashed line is used to separate members from candidates.
{\it Right:} Parallax distribution for the sample showing the observed values of $\varpi$ and respective error bars. Green $\blacklozenge$ symbols indicate the sources found in the secondary structure of the cloud (see Sect. \ref{sec:compara}). The vertical blue lines indicate the error-weighted mean parallax $\langle~\pi~\rangle$=0.91$\pm$0.02 mas (dashed lines show 3$\sigma$ deviation).
}
\label{fig:par}
\end{figure}

Table \ref{tab:pmov} gives the intervals of parameters adopted for the catalogue query, as well as the results found for our sample. The mean values obtained for proper motion define the 
membership criteria. 
An illustration of the reliability of the method is given in Fig. \ref{fig:par} (left panel), were the distribution of probabilities is presented as a function of modulus of proper motion
$| \mu |$ = $[ (\mu_\alpha)^2 + (\mu_\delta)^2]^{0.5}$.
We suggest that objects with membership probability P $\ge$ 50\% are very likely associated with the ambient cloud, which are considered here very-likely CMa members. 
These members have ($\mu_\alpha$, $\mu_\delta$) within 1$\sigma$ of the mean values found for the sample. 
Objects appearing below the dashed line (P$<$50\%) in Fig. \ref{fig:par} (left panel) are considered candidates or probable field-stars, since they have proper motion parameters in ranges that are
 larger than the 1$\sigma$ threshold adopted by us.

\begin{table*}
\begin{center}
\caption{Limits used for the {\it Gaia} data query and results from the kinematic members selection.}
\begin{tabular}{cccccc} 
\hline
ID & RA & Dec & $\varpi$ & $\mu_\alpha$ & $\mu_\delta$ \\
 & (J2000)&(J2000) & (mas) & (mas/yr) & (mas/yr) \\
\hline
query range & 07$^{\rm h}$13$^{\rm m}$ to 07$^{\rm h}$15$^{\rm m}$ & $-$12$^{\circ}$35$^{\prime}$ to $-$11$^{\rm o}$45$^{\prime}$ & 0.4 to 2 & $-$7 to 1 & $-$4 to 4 \\
P$\ge$50\% & 07$^{\rm h}$13$.\!\!^{\rm m}$7 to 07$^{\rm h}$14$.\!\!^{\rm m}$4& $-$12$^{\circ}$14$^{\prime}$ to $-$11$^{\rm o}$58$^{\prime}$ & 0.87 $\pm$ 0.30 & $-$3.26 $\pm$ 0.44 & 1.08 $\pm$ 0.46 \\
CMa06 & 07$^{\rm h}$03$.\!\!^{\rm m}$7 to 07$^{\rm h}$05$.\!\!^{\rm m}$6 & $-$11$^{\circ}$34$^{\prime}$ to $-$11$^{\rm o}$00$^{\prime}$ & 0.85 $\pm$ 0.09 & $-$4.18 $\pm$ 0.36 & 1.52 $\pm$ 0.21 \\
Sh~2-296 & 07$^{\rm h}$01$.\!\!^{\rm m}$2 to 07$^{\rm h}$06$.\!\!^{\rm m}$8 & $-$12$^{\circ}$12$^{\prime}$ to $-$10$^{\rm o}$48$^{\prime}$ & 0.8 to 1.25 & $-$4.10 $\pm$ 0.60 & 1.50 $\pm$ 0.40 \\
\hline
\end{tabular}
\label{tab:pmov}
\end{center}
\vspace{-0.2cm}
{\scriptsize Note: Our sample (P$\ge$50\%) is compared with clusters CMa06 \citep{SS20} and Sh~2-296 \citep{GH20} located to the W direction of the CMa shell.}
\end{table*}

\citet{SS20} use HDBScan technique to fit 5 parameters
 from {\it Gaia} aiming to explore the stellar clusters and sub-groups in the entire CMa region. No sub-group was found by them in the area studied here, probably
 due to the low number of members, and/or the objects are too faint to be identified by the automatic procedure. However, it is interesting to compare our results with those
 found by \citet{SS20} for the group they called as CMa06, which coincides with the Sh~2-296 nebula (see West cloud in Fig. \ref{fig:avmap}a).
Despite the fact that this cluster is located on the opposite side of the CMa shell ( $\sim$ 2$^{\rm o}$ to the W), its parallax and proper motion are quite similar to the results of our sample. This may be due to a common star formation history. A similar result was independently achieved by \citet[][]{GH20} in the study of objects associated to Sh~2-296 using the same method adopted here \citep{HGH19}, which validates our criteria to select the cloud members. The results found for CMa06 and Sh~2-296 are also presented in Table \ref{tab:pmov}.

In order to evaluate the error-weighted parallax, we adopted the calculation used by \citet{navarete19}, based on the uncertainty on measured parallax ($\varpi$) 
and the spatial correlation 
between the position of the sources. We obtained for our sample the mean parallax $\langle~\pi~\rangle$= 0.91$\pm$0.02 mas that was converted to the distance 
 of 1099$_{-24}^{+25}$ pc. Figure \ref{fig:par} (right panel) shows the distribution of parallaxes and error bars given in Table \ref{tab:members}, highlighting 11 stars
 (2 Class II and 9 Class III) that coincide with the secondary structure of the cloud found in the centre of the gas distribution, discussed in Sect. \ref{sec:compara}.
 
\subsection{Characterisation}

\subsubsection{Infrared excess}
\label{sec:wise}

The infrared (IR) data from {\it AllWISE} catalogue were used for two purposes: (i) characterizing the stars that were selected as members of CMa, on the basis of 
proper motion, and (ii) searching for embedded sources that were not detected by {\it Gaia} probably due to high levels of extinction in dense regions. The query was restricted 
to the same area described in Sect. \ref{sec:gaia}. 

Our analysis of IR-excess is based on colour-colour diagrams using WISE bands: W1 (3.4 $\mu$m), W2 (4.6 $\mu$m) and W3 (12 $\mu$m) that are useful to 
distinguish different classes of pre-main sequence stars. Based on the results from \citet{Rebull14} for the Taurus star-forming region,
 \citet{K14} proposed the limits on the [W1$-$W2] $\times$ [W2$-$W3] diagram defining the expected locus for Class I and Class II objects, due to their significant IR-excess 
 compared with Class III objects and field-stars.

In order to ensure the photometric quality, we extracted from the {\it AllWISE} catalogue only the sources in agreement with the following conditions for the magnitude 
measured at 12 $\mu$m: $0.45 < W3_{r\chi^2} <1.15$ and $W3_{snr} > 5$, where $r\chi^2$ and $snr$ correspond to the photometric error and signal-to-noise ratio, respectively. According to \citet{K14} these filters are applied with the purpose of mitigating contamination from fake detections. 

\begin{figure}
\begin{center}
\includegraphics[width=0.5\textwidth]{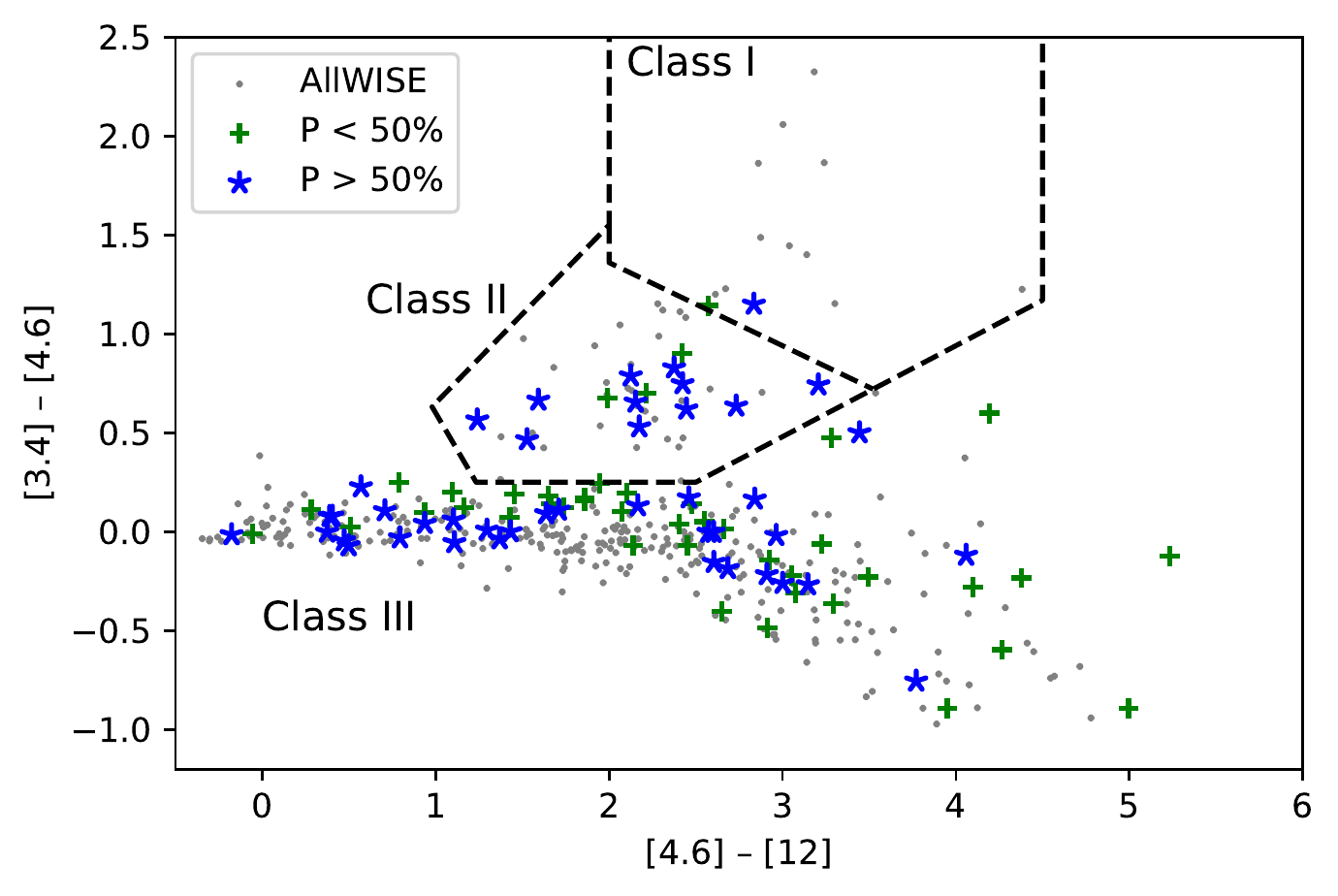} 
\end{center}
\caption{WISE colour-colour diagram displaying the expected locus for Class I and Class II sources as proposed by \citet{K14}. 
Confirmed members (blue $\star$) have P $\ge$ 50\%, while possible members (green $+$) have
 P $<$ 50\%. The remaining {\it AllWISE} objects (grey dots) are not present in {\it Gaia} EDR3 catalogue. IR sources found in the Classes I and II boxes are considered here as additional candidates.
}
\label{fig:wisebox}
\end{figure}

Figure \ref{fig:wisebox} displays the [W1$-$W2] $\times$ [W2$-$W3] diagram for 383 sources that coincide with the 
fields observed with IRAM-30m and that are common on both catalogues: {\it Gaia} and WISE, which membership 
probability is represented by different symbols. It can be noted that most of the studied objects (89\%) are plotted in the region where Class III and/or field-stars are
expected to be found. 

Our final sample contains 40 members (1 is Class I; 11 are Class II; and 28 are Class III) confirmed by proper motion (P$\ge$ 50\%), and 4 candidates (1 is Class I, and 3 are Class II). Since we are mainly interested in embedded objects, Class III candidates (P$<$50\%) were not included in the sample. Table \ref{tab:members} gives the list of objects and the {\it Gaia EDR3} parameters used in the membership probability calculation. 
The division between the classes of objects is indicated by double-lines in Table \ref{tab:members}.

Aiming to complement the list of objects, we also searched for WISE sources not detected by {\it Gaia}. 
Due to the lack of membership information for these additional sources, we consider here as possible members only Class I or Class II objects. In other words, among the 
WISE candidates not detected by {\it Gaia}, the Class III objects were not taken into account due to the difficult in distinguishing them from field-stars. By this way, the sample is complemented by 45 WISE 
sources that we consider possible members (12 are Class I, and 33 are Class II). The distribution of the sources in the equatorial coordinates space
 is shown in Fig. \ref{fig:radec} (left panel). The infrared photometry used in the analysis of the stellar population is given in Table \ref{tab:IR} for the same list presented in Table \ref{tab:members}.

We have verified in our sample the presence of H$\alpha$ emitters by using a cross-correlation with the results from \citet{peter} that revealed 353 new H$\alpha$ stars in the direction of CMa. The area surveyed by us contains 10 H$\alpha$ stars. However, only 4 of them coincide with the
CMa members, meaning that the other H$\alpha$ stars have proper motion and/or parallax in disagreement with the kinematic selection criteria adopted by us. In Table \ref{tab:members}
 we add a comment identifying each of the H$\alpha$ stars identified by \citet{peter}, which are classified by us as Class II objects. For classical T Tauri stars, the H$\alpha$ emission is an evidence of accretion process that is expected for Class II objects and is also related to the presence of a circumstellar disk. The characteristics of these 4 H$\alpha$ stars are in agreement with their young age ($<$ 5 Myr), which is estimated in the next section.

\subsubsection{Cluster Age}
\label{sec:age}

In order to confirm the youth of the sample we constructed the color-magnitude diagram using the photometric data at bands $G$ ($\sim$ 600 nm), 
G$_{BP}$ ($\sim$ 500 nm), and G$_{RP}$ ($\sim$ 700 nm) from {\it Gaia} EDR3. The magnitudes were corrected for reddening by adopting the 
A$_{\lambda}$/A$_V$ relations from \citet{Cardelli01}. For each source, we calculate the distance modulus given by its parallax, which was used to estimate the unreddened absolute magnitude $(M_G)_0$.
 
Figure \ref{fig:radec} (right panel) shows the distribution of our sample in the $(M_G)_0 \times [G_{BP} $-$ G_{RP}]_0$ diagram compared with isochrones from 
{\it PARSEC} {\footnote{Version v1.2S+COLIBRI PR16 of {\it PARSEC} models available on http://stev.oapd.inaf.it/cgi-bin/cmd.}} \citep{Bressan12, Marigo17}. 
The models for 1 Myr and 5 Myr were adopted to indicate
the range of ages previously reported for the CMa young population \citep{GH09}, while the 100 Myr isochrone was chosen as representative of the ZAMS ({\textit Zero Age Main Sequence}).

In the color-magnitude diagram we display only the sources with good photometric data, which show snr~$>$~10 in all bands. Some of the Class III objects show a good fit with the ZAMS indicating that we adopted a suitable mean value of visual extinction (A$_V$ = 0.9 mag) in the reddening correction. 
Since the Class III objects are not affected by circumstellar IR-excess, we argue that
this low level of A$_V$ corresponds to the interstellar reddening in the direction of CMa, which is in agreement with the extinction map of the region \citep{GH08}. 
We conclude that it is highly probable that the Class III stars are observed in the near vicinity, but are preferentially located in the foreground, of the cloud.

The same can be said for other sources detected by {\it Gaia}, since they are visible. However, some of them are too faint ($(M_G)_0 >$ 8 mag in right panel of Fig. \ref{fig:radec}), probably due to a reddening that needs be corrected with a higher value of A$_V$, which should be evaluated individually. This can be the case of sources that are still surrounded by some amount of cloud material. 

We are aware that using optical color-magnitude diagram gives only a rough estimation of age. Despite of that, it can be noted that most of the sources exhibiting IR excess (Class I and Class II) are $\sim$ 5 Myr or younger. 
Several of the Class III presented here are in the same range of age and a few sources seem to be older, but still are in the pre-main sequence phase, confirming the youth of our sample. Among the {\it Gaia} EDR3 sources, we do not find massive stars. The brightest Class III objects have colors similar to 2 M$_{\odot}$ stars. 

As mentioned in Sect. \ref{sec:gaia} the proper motion of our sample, which is located in the E side of the CMa shell, coincides with the values found for group CMa06 located to the W. The estimated age for this group is 6$\pm$1 Myr \citep{SS20}, suggesting that these two stellar clusters, which are located at opposite sides, may have had a 
similar star formation scenario.


\section{Comparing gas and star distributions}
\label{sec:compara}

\begin{figure}
\begin{center}
\includegraphics[width=0.24\textwidth]{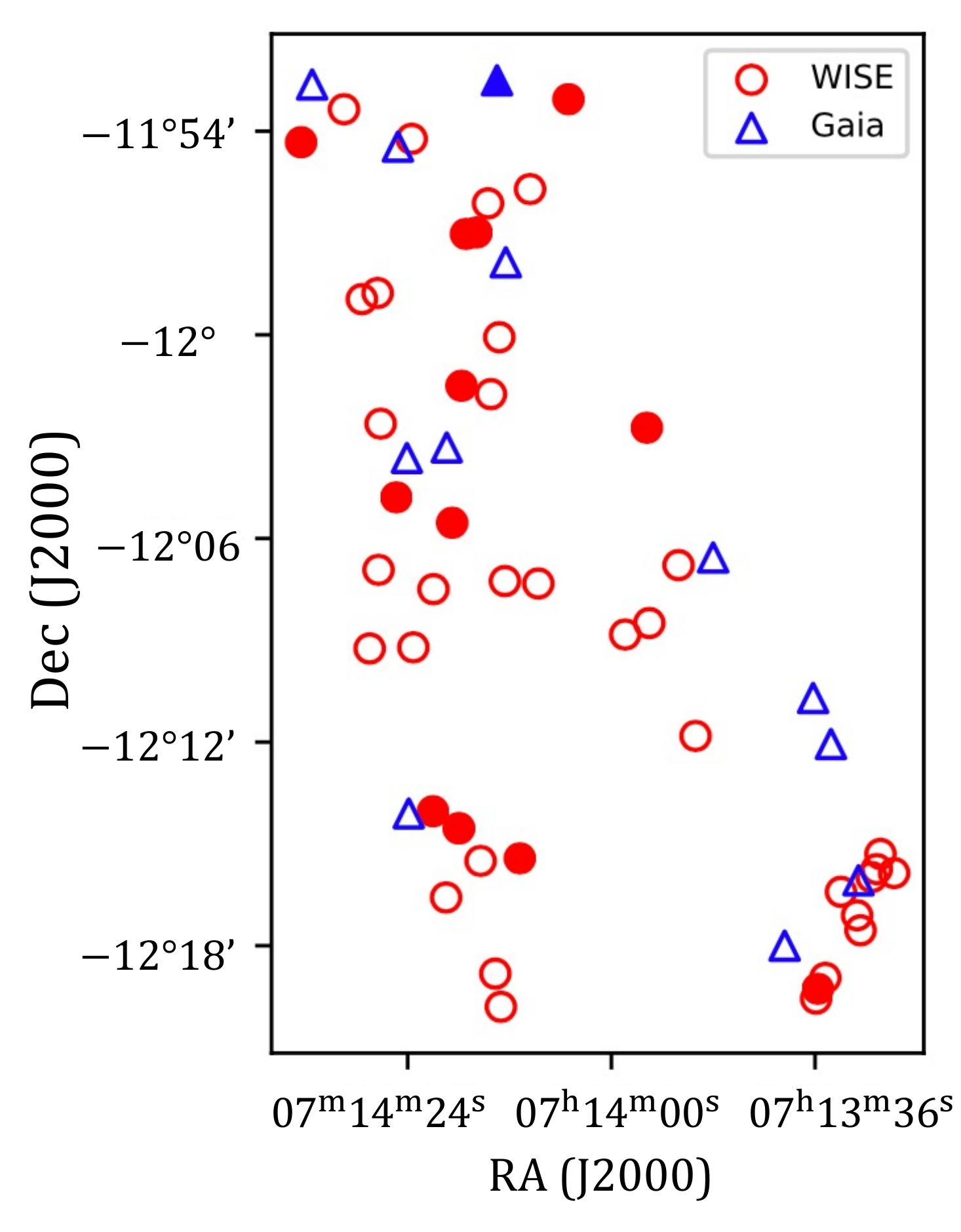} 
\includegraphics[width=0.24\textwidth]{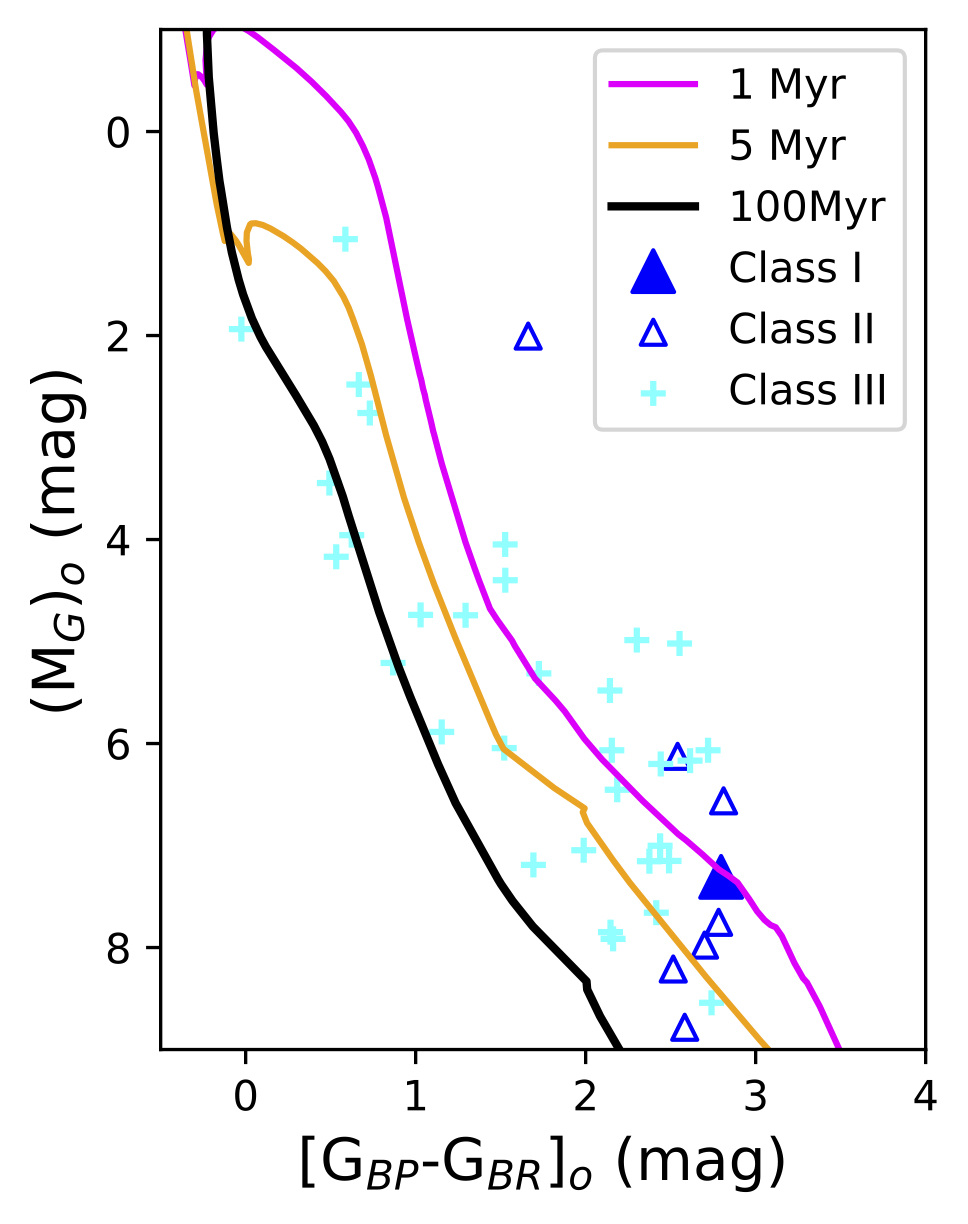} 
\end{center}
\caption{{\it Left:} Equatorial coordinates of Class I (filled symbols) and Class II (open symbols) that are CMa members (triangles) and additional candidates from {\it AllWISE} catalogue (circles). 
 {\it Right:} Absolute magnitude as a function of colour from {\it Gaia} data for the CMa members (P $\ge$ 50\%). The magnitudes are corrected for extinction. We use isochrones from PARSEC models, by adopting the 100 Myr line as representative of the ZAMS. }
\label{fig:radec}
\end{figure}

As can be seen in Figs.~\ref{fig:spire} and \ref{fig:radec} the projected spatial distribution of our sample of stars closely follows the filamentary structure present in the $^{13}$CO gas. Most of the Class II sources are found around dense cores, probably emerged from the cloud and still associated with it.

\begin{figure}
\begin{center}
\includegraphics[width=9cm]{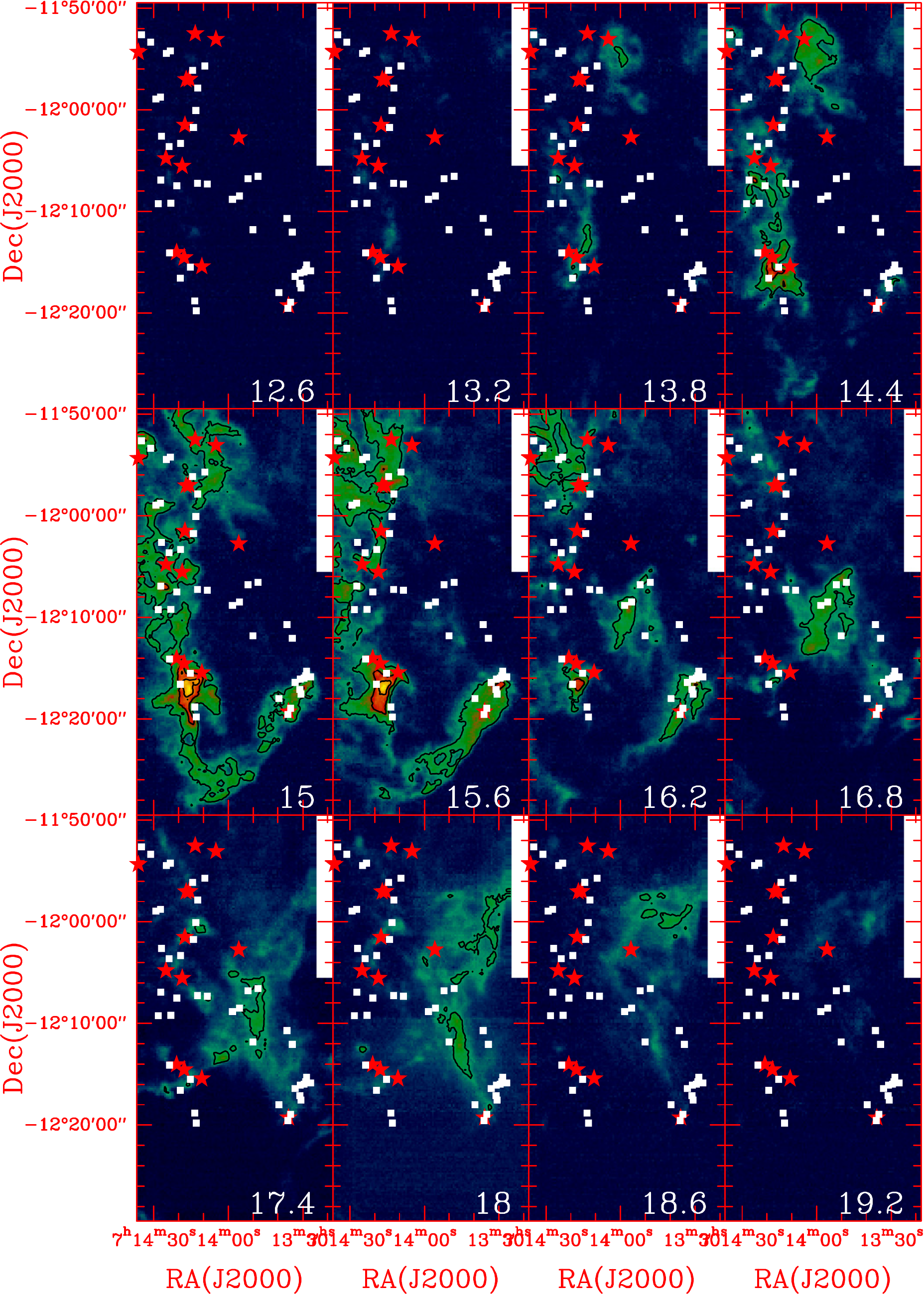} 
\end{center}
\caption{Molecular gas mapping ($^{13}$CO) as a function of velocity given in km/s (see right bottom of each panel). The position of WISE sources is shown for Class I (red stars) and Class II (white squares) objects.}
\label{fig:molgas}
\end{figure}

 In Fig.~\ref{fig:molgas} we plot a mosaic of $^{13}$CO maps showing in each panel a different range of gas velocity, from 12.6 km/s to 19.2 km/s. A main structure with V = 15 $\pm$ 1.2 km/s is clearly seen at 
 RA$\sim$ 07$^{\rm h}$14$.\!\!^{\rm m}$5, growing from N to S and then continuing to SW. A secondary structure at V = 17.4 $\pm$ 1.2 km/s appears then at the centre of the plot (RA = 07$^{\rm h}$14$^{\rm m}$, Dec = 12$^{\circ}$10$^{\prime}$). Despite their partial overlapping, they seem to be two structures representing different parts and/or movements of the cloud. 
The relative motions of these two gas structures are suggestive of a large-scale expanding gas motions around the cavity detected in the molecular gas at IRAM and in the continuum emission with SPIRE (see Fig. \ref{fig:spire}). It is remarkable that the ``cavity" looks void of young stars and protostars. 

\begin{figure}
\begin{center}
\includegraphics[width=0.48\textwidth]{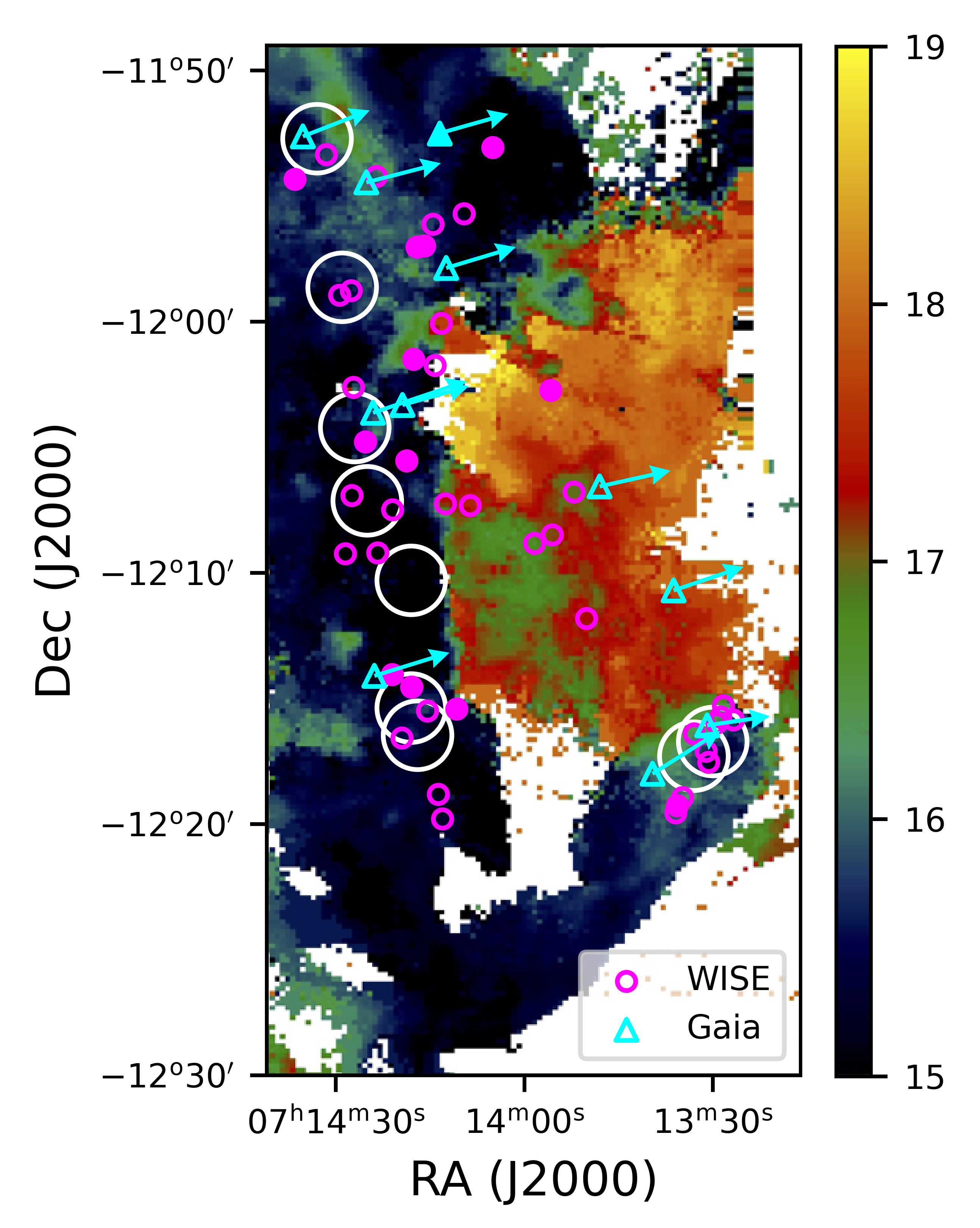} 
\end{center}
\caption{Moment 1 map showing the distribution of velocities measured in km/s, based on $^{13}$CO data. The proper motion vectors
are displayed for the YSOs that are Gaia sources (same symbols as Fig. \ref{fig:radec}). White circles indicate the position of dense cores (see Fig. \ref{fig:spire}).}
\label{fig:sbm}
\end{figure}

The presence of these two cloud structures and their connection with the stellar population is confirmed by analyzing the gas kinematics through moment maps similar to the works by \citet{A21} and \citet{S16}, for instance. Based on the $^{13}$CO observations, we obtained the Moment 1 (mean velocities) map that is presented in Fig. \ref{fig:sbm} overlaid by the distribution of the YSOs that are Class I and Class II objects (the same presented in Fig. \ref{fig:radec}). For Gaia sources we also display the proper motion vectors, in order to compare the velocity dispersion of stars with the radial velocity of the gas. The map clearly shows the secondary structure with velocities V $>$ 17 km/s, while the gas in the main structure of the cloud has velocities ranging from less than 15~km/s to $\sim$ 16.5~km/s, which supports the discussion based on Fig. \ref{fig:molgas}. For our sample of YSOs, the proper motion velocity dispersion, measured in the plane of the sky, shows no remarkable trend that could indicate non-isotropic velocity patterns. This prevents us to address here a deeper integration of the stellar content with the gas kinematics for cloud sub-structures.

Mapping of the molecular line emission reveals the presence of dense cores, detected in the lines of $^{13}$CO and CS. A sample of 9 cores was investigated in more detail. The $^{12}$CO and $^{13}$CO line profiles are displayed in the right panel of Fig. \ref{fig:spire}. The observed line profiles were fitted by using a simple modeling of the$^{13}$CO and C$^{18}$O lines,
adopting a canonical relative abundance ratio [$^{13}$CO]/[C$^{18}$O] = 8. This
allowed us to estimate the optical depths $\tau_{13}$ and $\tau_{18}$ of
both isotopologues. The observed line profiles and optical depths were
subsequently modelled with the radiative transfer code MADEX \citep{Cernicharo12} in the Large-Velocity Gradient approximation \citep[][]{Sobolev58,Sobolev60} 
using the CO-H$_2$ collisional coefficients of \citet{Yang} and the line-width (FWHM) measured from a gaussian fit to the line
profiles. We could check that the results are essentially independent of
n(H$_2$) in the regime of densities 10$^4$ $-$ 10$^6$ cm$^{-3}$, typical of dark clouds and
star-forming cores, i.e the ground state transitions are thermalized.

 MADEX was adopted by us to derive the kinetic temperature and the molecular gas column density in the cores. The main result is that these cores consist of cold gas, with temperatures in the range 8--13~K. One of the most direct signs of protostellar activity is the signature of mass-loss phenomena (outflows) as broad wings in the CO rotational line profiles observed towards dense cores. This is illustrated in the CO lines profiles of cores 1, 3, 4, 5, 6 and 7 in Fig. \ref{fig:spire}. In the case of core 4, the outflow wings reach velocities as high as 20 km/s that is clear evidence for active, possibly massive, star formation in the core. Some of these cores, like e.g. cores 6 and 7, are associated with IR ({\it WISE}) sources, and we note that many of the IR sources coincide with the $^{13}$CO cores identified in our map. We speculate that the absence of IR sources in the other cores which display bipolar outflow signatures indicates that the driving protostars might be at earlier stage of evolution, possibly Class~0. 

Conversely, other dense cores detected in our observations appear to be in a quiescent stage, as illustrated by cores 2, 8 and 9 (Fig. \ref{fig:spire}). Based on our molecular gas content analysis presented in Table \ref{tab:coeurs}, it appears that the cores in the Northern part of the region ($n \geq 6$) tend to have lower gas column densities. More work is needed to confirm this trend (higher densities found in the Southern part of the cloud), based on a very reduced sample, and to investigate whether it is related to the physical or chemical (depletion ?) evolution of the region. 
 
Despite all the dense cores (kinetic temperatures from 10 to 14 K) are found in the main structure, with most of the stellar members associated to it, there are 1 Class I and 8 Class II sources located in the direction of the secondary, smaller cloud structure. Since only two of these stars are {\it Gaia} sources, we cannot evaluate kinematic differences compared with the stars associated with the main structure. 

Indeed, no trend is observed when comparing the spatial distribution of sources having proper motion vectors displayed in Fig. \ref{fig:sbm}. The same can be said for the parallax distribution (see Fig. \ref{fig:par}), where 9 Class III sources are also highlighted because they appear in the secondary structure, but without showing trends on $\varpi$. 
However, it is interesting to note that 2MASS07134800-1206336, found in the secondary structure, is the brightest star of the Class II objects (G$_{Gaia}$ = 13 mag), which appears 
in the red side of the color-magnitude diagram (see Fig. \ref{fig:radec}, right panel), with age $<$ 1 Myr. The parallax of this bright source
indicates it is in the largest distance ($\varpi$ = 0.62 $\pm$ 0.04) compared with the other members. Therefore, it is possible that the reddening still needs to be corrected for this source and its mass shall be larger than 2 M$_{\odot}$.

\begin{table*}
\begin{center}
\caption{Membership probability, parallax, proper motion and {\it Gaia} EDR3 photometry.}
\small
\begin{tabular}{cccccccc} \hline{\it Gaia} EDR3 & P & $\varpi$ & $\mu_\alpha$ & $\mu_\delta$ & G & G$_{BP}$ & G$_{RP}$ \\ 
 & \% & mas & mas/yr & mas/yr & mag & mag & mag \\ 
\hline
3045239083971370112 & 87 & $0.77\pm0.17$ & $-3.21\pm0.16$ & $0.91\pm0.15$ & $18.65$ & $20.43$ & $17.24 $ \\ 
3045208886056506112 & 13 & $1.69\pm1.10$ & $-2.52\pm0.86$ & $0.97\pm1.06$ & $20.33$ & $21.63$ & $18.78 $ \\ 
\hline \hline
3045225516172964864$^a$ & 95 & $0.66\pm0.02$ & $-3.32\pm0.02$ & $0.74\pm0.02$ & $13.80$ & $14.77$ & $12.74 $ \\ 
3045229291445342592$^b$ & 92 & $1.31\pm0.19$ & $-3.26\pm0.18$ & $0.98\pm0.16$ & $18.80$ & $20.42$ & $17.43 $ \\ 
3045227749555916032$^c$ & 91 & $0.61\pm0.14$ & $-3.11\pm0.16$ & $0.88\pm0.13$ & $17.93$ & $19.59$ & $16.63 $ \\ 
3045131301773079168 & 90 & $0.79\pm0.10$ & $-3.02\pm0.08$ & $1.04\pm0.09$ & $17.44$ & $19.06$ & $16.22 $ \\ 
3045208950479742208 & 83 & $1.12\pm0.26$ & $-3.28\pm0.20$ & $1.11\pm0.18$ & $18.89$ & $20.72$ & $17.57 $ \\ 
3045237434703416320 & 82 & $1.63\pm0.39$ & $-3.52\pm0.34$ & $1.09\pm0.33$ & $19.89$ & $21.26$ & $18.52 $ \\ 
3045235785435946624 & 79 & $0.72\pm0.32$ & $-3.13\pm0.26$ & $1.22\pm0.24$ & $19.03$ & $20.96$ & $17.65 $ \\ 
3045117141257872000 & 78 & $1.00\pm0.30$ & $-3.49\pm0.28$ & $0.90\pm0.29$ & $19.01$ & $20.52$ & $17.75 $ \\ 
3045215998525392384$^d$ & 65 & $0.97\pm0.11$ & $-2.93\pm0.11$ & $0.43\pm0.10$ & $17.29$ & $19.13$ & $15.91 $ \\ 
3045116114758900736 & 65 & $0.57\pm0.52$ & $-4.38\pm0.53$ & $1.48\pm0.54$ & $19.99$ & $21.22$ & $18.57 $ \\ 
3045114804796665984 & 63 & $1.02\pm0.51$ & $-3.17\pm0.44$ & $2.00\pm0.48$ & $19.85$ & $21.54$ & $18.39 $ \\ 
3045214070080777600 & 27 & $0.92\pm1.18$ & $-2.78\pm0.90$ & $0.44\pm1.00$ & $20.66$ & $21.37$ & $19.22 $ \\ 
3045234170527254656 & 21 & $0.45\pm0.10$ & $-3.00\pm0.10$ & $0.87\pm0.09$ & $17.39$ & $19.19$ & $16.07 $ \\ 
3045235785436006400 & 21 & $1.47\pm0.80$ & $-3.48\pm0.62$ & $0.54\pm0.63$ & $20.26$ & $21.51$ & $18.81 $ \\ 
\hline \hline
3045118416872687744 & 92 & $0.96\pm0.03$ & $-3.69\pm0.03$ & $0.99\pm0.03$ & $12.85$ & $12.97$ & $12.63 $ \\ 
3045226821842999552 & 97 & $0.80\pm0.05$ & $-3.51\pm0.06$ & $1.03\pm0.06$ & $16.32$ & $17.39$ & $15.29 $ \\ 
3045213352825581312 & 97 & $0.89\pm0.02$ & $-3.28\pm0.02$ & $0.99\pm0.02$ & $14.03$ & $14.49$ & $13.40 $ \\ 
3045228368031224448 & 96 & $0.84\pm0.05$ & $-3.23\pm0.05$ & $1.15\pm0.05$ & $15.86$ & $16.64$ & $14.99 $ \\ 
3045215758007221632 & 94 & $0.79\pm0.04$ & $-3.19\pm0.04$ & $0.94\pm0.04$ & $15.85$ & $16.76$ & $14.87 $ \\ 
3045232555619709056 & 93 & $1.17\pm0.02$ & $-3.60\pm0.02$ & $0.83\pm0.02$ & $14.50$ & $14.87$ & $13.97 $ \\ 
3045114809098665984 & 92 & $0.76\pm0.02$ & $-3.22\pm0.01$ & $1.01\pm0.01$ & $12.60$ & $12.99$ & $12.02 $ \\ 
3045229291444432384 & 90 & $0.50\pm0.18$ & $-4.08\pm0.16$ & $1.06\pm0.14$ & $18.59$ & $20.14$ & $17.32 $ \\ 
3045114701716408832 & 90 & $0.90\pm0.21$ & $-3.53\pm0.18$ & $1.11\pm0.20$ & $18.70$ & $20.30$ & $17.46 $ \\ 
3045114843459884288 & 90 & $0.87\pm0.41$ & $-3.71\pm0.33$ & $1.46\pm0.36$ & $19.60$ & $20.97$ & $18.36 $ \\ 
3045209156636885760 & 85 & $0.72\pm0.13$ & $-2.71\pm0.11$ & $0.94\pm0.09$ & $17.68$ & $19.11$ & $16.52 $ \\ 
3045114396781820672 & 84 & $0.91\pm0.03$ & $-3.11\pm0.02$ & $0.96\pm0.02$ & $14.53$ & $14.88$ & $14.02 $ \\ 
3045118382511875712 & 82 & $0.88\pm0.05$ & $-3.90\pm0.04$ & $0.97\pm0.04$ & $15.81$ & $16.43$ & $15.05 $ \\ 
3045227509039371648 & 80 & $0.44\pm0.37$ & $-2.95\pm0.36$ & $0.94\pm0.30$ & $19.39$ & $20.99$ & $18.11 $ \\ 
3045233242814458752 & 79 & $0.67\pm0.19$ & $-2.92\pm0.20$ & $1.19\pm0.22$ & $18.63$ & $20.13$ & $17.47 $ \\ 
3045229879859736704 & 70 & $0.51\pm0.10$ & $-4.33\pm0.09$ & $1.11\pm0.10$ & $17.53$ & $18.91$ & $16.38 $ \\ 
3045207885326442496 & 70 & $1.27\pm0.31$ & $-3.06\pm0.24$ & $0.87\pm0.23$ & $19.34$ & $20.71$ & $18.19 $ \\ 
3045228952146750208 & 70 & $0.55\pm0.13$ & $-2.88\pm0.14$ & $1.02\pm0.12$ & $18.11$ & $20.01$ & $16.81 $ \\ 
3045130678995758848 & 69 & $1.37\pm0.28$ & $-3.63\pm0.33$ & $1.67\pm0.34$ & $19.26$ & $20.84$ & $17.97 $ \\ 
3045130472837058176 & 66 & $1.56\pm0.59$ & $-2.26\pm0.70$ & $0.83\pm0.72$ & $20.20$ & $21.74$ & $18.88 $ \\ 
3045420954362862336 & 64 & $1.38\pm0.03$ & $-3.52\pm0.03$ & $1.56\pm0.03$ & $15.32$ & $15.85$ & $14.62 $ \\ 
3045212936208209024 & 64 & $0.60\pm0.39$ & $-2.34\pm0.30$ & $1.02\pm0.33$ & $19.33$ & $20.77$ & $18.11 $ \\ 
3045233693790633472 & 62 & $1.25\pm0.19$ & $-2.80\pm0.17$ & $1.70\pm0.16$ & $18.58$ & $19.99$ & $17.43 $ \\ 
3045233281473783936 & 59 & $0.92\pm0.01$ & $-3.01\pm0.02$ & $1.00\pm0.02$ & $13.50$ & $13.92$ & $12.89 $ \\ 
3045213623404184448 & 58 & $1.37\pm0.40$ & $-2.63\pm0.36$ & $0.66\pm0.31$ & $19.46$ & $21.21$ & $18.14 $ \\ 
3045233998728534272 & 56 & $0.56\pm0.37$ & $-2.45\pm0.29$ & $1.47\pm0.27$ & $19.58$ & $20.97$ & $18.31 $ \\ 
3045233487625692160 & 53 & $0.81\pm0.57$ & $-3.79\pm0.78$ & $1.65\pm0.95$ & $20.07$ & $21.24$ & $18.75 $ \\ 
3045225202636012800 & 50 & $1.06\pm0.58$ & $-3.88\pm0.61$ & $0.45\pm0.64$ & $20.04$ & $21.43$ & $18.73 $ \\ 
\hline
\end{tabular}
\label{tab:members}
\end{center} 
\vspace{-0.3cm}
{\small Notes: The objects are separated by: Class I (top); Class II (middle); Class III (bottom), according with Table \ref{tab:IR}.
Stars identified as H$\alpha$ emitters are indicated by (a) ID~381; (b) ID~387; (c) ID~390; and (d) ID~392, where the ID numbers are given by \citet{peter}.}
\end{table*}

 
\section{Conclusions}
\label{sec:conclude}

We mapped molecular clouds with the IRAM-30m to investigate the properties of cores and filaments in order to determine the 
spatial and kinematic distributions of gas in the cloud and their relation with the star formation activity. In the general context, 
we aim 
to search for and characterise the shock driven by the SN into the molecular cloud, which would be responsible for the gas compression and gravitational collapse. 

Comparing the spatial distribution of the stellar population with the cores revealed by the $^{13}$CO map, we verify that peaks of emission coincide with the position of YSOs (see Fig. \ref{fig:spire}). 
These results confirm that CMa harbors pre- and protostellar cores showing that star formation is still underway in cores distributed along filaments,
as suggested by \citet{Elia13} based on a survey with {\it Herschel}.

We selected a sample of 89 sources, 40 of them are confirmed members ({\it Gaia} EDR3) associated to CMa.
The mean error-weighted parallax was converted to the distance 
 of 1099$_{-24}^{+25}$ pc, which is in excellent agreement with previous results for the CMa region \citep{Claria74,Tov93,Shev99}. 
Using the WISE colors, the sources were characterized according with the IR-excess. We are mainly interested here on Classes I and II that may have circumstellar disk. 
All of these disk-bearing stars are found around the filamentary structure of the cloud, \break
despite several of them are not embedded, but probably located in the cloud foreground.
The color-magnitude diagram constructed with the G$_{Gaia}$ photometry were used to verify whether the selected sources truly are pre-main sequence stars. The ages
are less than 5 Myr for most of the sources, coinciding with ages estimated for the cluster found in the opposite side of the CMa shell.
 Moreover, as shown in Fig. \ref{fig:wisebox} (right panel) and discussed below, the presence of embedded objects corresponding to protostellar phase (Class 0 and Class I sources) is confirmed in our sample. Since Class 0 and Class I stars have typical ages of 10$^4$ to 10$^5$ yr, they can be considered a direct evidence for sequential star formation in the region.
 
The kinematic analysis reveals the presence of two structures representing different parts of the cloud.
 A main structure, exhibiting V = 15 $\pm$ 1.2 km/s, is formed from N to S in an extended stripe where most of the dense cores are found and YSOs are associated with.
The secondary structure, seen at V = 17.4 $\pm$ 1.2 km/s, grows from the SW to the centre of Fig. \ref{fig:spire} (middle panel).
Evidence of outflows is remarkable from the broad wings in the CO rotational line profiles observed in 6 dense cores, which are suggested to be sites of protostellar activity as indicated by their association with IR sources,
and shows that a burst of star formation is currently going on. 
Follow up observations, such as near-IR spectroscopy to obtain radial velocity of the members, are required to 
confirm the trends of stellar members following the cloud structure observed for this limited sample of objects associated to the studied region. 

Despite these are partial results, they shed some light attempting to explain the star formation in CMa. The age of the Class II population is consistent with the age/timescale of the SNRs \citep{F19}. The similarity of the young stellar population properties on East/Western sides of the shell supports the scenario of a large-scale gravitational collapse of the shell, most likely induced by the SN explosions. 

We tentatively suggest that the presence of young protostellar population (with massive and/or intermediate-mass objects, as indicated by the strong outflow emission) could indicate that a second episode of star formation has recently taken place. 
The distribution of YSOs in the two gas layers at 15 and 17 km/s suggests that local events on the Eastern side (the region surveyed at IRAM) have contributed to drive the current star formation process. 
Our observations of the region altogether support the large-scale scenario of SNR shock-driven, but other processes at smaller scales appear to have influenced and possibly induced the local star formation.

\begin{table*}
\begin{center}
\small
\caption{Infrared photometry for stellar population.}
\begin{tabular}{cccccccccc} \hline
{\it Gaia} EDR3 & 2MASS & J & H & K & W1 & W2 & W3 & W4 & Class \\
 & & mag & mag & mag & mag & mag & mag & mag &\\
\hline 
3045239083971370112 & 07141347-1152298 & 14.44 & 12.88 & 11.82 & 10.97 & 9.82 & 6.99 & 4.61 & I \\ 
3045208886056506112 & 07141796-1214324 & 15.55 & 14.05 & 13.13 & 11.60 & 10.46 & 7.88 & 5.63 & I \\ 
\hline \hline
3045225516172964864 & 07134800-1206336 & 10.79 & 9.62 & 8.65 & 7.62 & 7.00 & 4.55 & 2.66 & II \\ 
3045229291445342592 & 07141244-1157517 & 15.24 & 13.89 & 12.94 & 11.40 & 10.61 & 8.49 & 6.41 & II \\ 
3045227749555916032 & 07141939-1203191 & 13.39 & 12.11 & 11.29 & 10.32 & 9.66 & 7.51 & 5.74 & II \\ 
3045131301773079168 & 07133622-1210420 & 14.09 & 12.89 & 12.24 & 12.04 & 11.40 & 8.67 & 6.45 & II \\ 
3045208950479742208 & 07142387-1214061 & 15.06 & 13.94 & 13.45 & 12.66 & 11.91 & 8.71 & 5.69 & II \\ 
3045237434703416320 & 07143525-1152374 & 15.79 & 14.37 & 13.56 & 12.79 & 12.13 & 10.54 & 5.78 & II \\ 
3045235785435946624 & 07142515-1154273 & 15.02 & 13.74 & 12.92 & 11.77 & 11.20 & 9.96 & 8.38 & II \\ 
3045117141257872000 & 07133088-1216043 & 15.21 & 14.02 & 13.45 & 12.73 & 11.98 & 9.56 & 7.42 & II \\ 
3045215998525392384 & 07142409-1203377 & 13.20 & 11.99 & 11.32 & 10.40 & 9.57 & 7.20 & 4.77 & II \\ 
3045116114758900736 & 07133958-1217599 & 15.54 & 14.04 & 13.27 & 12.56 & 12.03 & 9.86 & 7.85 & II \\ 
3045114804796665984 & 07140707-1217153 & 15.73 & 14.87 & 14.32 & 13.98 & 13.48 & 10.03 & 7.68 & II \\ 
3045214070080777600 & 07142847-1209146 & 15.65 & 14.24 & 13.20 & 12.39 & 11.49 & 9.07 & 6.47 & II \\ 
3045234170527254656 & 07142755-1158460 & 13.72 & 12.58 & 12.07 & 11.39 & 10.69 & 8.47 & 6.48 & II \\ 
3045235785436006400 & 07142353-1154135 & 15.57 & 14.19 & 13.32 & 12.39 & 11.71 & 9.72 & 7.41 & II \\ 
\hline \hline
3045118416872687744 & 07133684-1216484 & 12.36 & 12.30 & 12.23 & 12.20 & 12.22 & 12.39 & 9.09 & III \\ 
3045226821842999552 & 07135758-1203138 & 13.71 & 12.96 & 12.79 & 12.76 & 12.68 & 12.27 & 9.14 & III \\ 
3045213352825581312 & 07140300-1208579 & 12.61 & 12.20 & 12.07 & 12.04 & 12.05 & 11.67 & 8.50 & III \\ 
3045228368031224448 & 07135653-1202357 & 13.74 & 13.14 & 12.92 & 12.84 & 12.88 & 12.41 & 9.07 & III \\ 
3045215758007221632 & 07143221-1204404 & 13.33 & 12.83 & 12.50 & 12.47 & 12.39 & 12.00 & 8.71 & III \\ 
3045232555619709056 & 07140255-1156164 & 13.36 & 13.02 & 12.93 & 12.95 & 12.98 & 11.62 & 8.90 & III \\ 
3045114809098665984 & 07141006-1217243 & 11.28 & 11.12 & 11.00 & 10.87 & 10.87 & 9.43 & 7.40 & III \\ 
3045229291444432384 & 07141365-1158183 & 15.07 & 14.21 & 13.78 & 13.53 & 13.47 & 12.37 & 9.00 & III \\ 
3045114701716408832 & 07141366-1217431 & 15.13 & 14.28 & 13.94 & 13.62 & 13.63 & 10.67 & 8.42 & III \\ 
3045114843459884288 & 07140167-1217317 & 16.24 & 15.64 & 15.33 & 15.60 & 15.44 & 12.60 & 9.21 & III \\ 
3045209156636885760 & 07143327-1213273 & 14.53 & 13.68 & 13.34 & 13.10 & 13.05 & 12.12 & 9.17 & III \\ 
3045114396781820672 & 07140309-1218168 & 13.47 & 13.14 & 13.07 & 13.03 & 13.08 & 11.97 & 8.84 & III \\ 
3045118382511875712 & 07134177-1216279 & 14.08 & 13.51 & 13.40 & 13.28 & 13.31 & 12.52 & 8.74 & III \\ 
3045227509039371648 & 07140828-1203310 & 15.77 & 14.66 & 14.25 & 14.07 & 13.98 & 12.34 & 8.96 & III \\ 
3045233242814458752 & 07134763-1156272 & 15.44 & 14.65 & 14.34 & 14.28 & 14.44 & 11.83 & 9.05 & III \\ 
3045229879859736704 & 07132666-1202354 & 14.46 & 13.69 & 13.38 & 13.20 & 13.19 & 11.89 & 8.18 & III \\ 
3045207885326442496 & 07142893-1216095 & 16.30 & 15.84 & 15.17 & 15.37 & 16.12 & 12.35 & 9.05 & III \\ 
3045228952146750208 & 07141111-1159429 & 14.36 & 13.45 & 12.94 & 12.70 & 12.59 & 11.88 & 9.09 & III \\ 
3045130678995758848 & 07133413-1212033 & 15.80 & 14.49 & 13.90 & 13.19 & 12.72 & 11.19 & 8.08 & III \\ 
3045130472837058176 & 07133056-1214195 & 16.44 & 15.26 & 15.36 & 14.78 & 14.65 & 12.48 & 8.72 & III \\ 
3045420954362862336 & 07133375-1153320 & 13.75 & 13.22 & 13.03 & 13.10 & 13.17 & 12.67 & 8.95 & III \\ 
3045212936208209024 & 07140301-1211153 & 16.01 & 15.16 & 14.53 & 14.51 & 14.40 & 12.69 & 9.17 & III \\ 
3045233693790633472 & 07135366-1153340 & 15.90 & 15.21 & 14.91 & 14.88 & 15.07 & 12.38 & 9.05 & III \\ 
3045233281473783936 & 07135236-1156044 & 12.15 & 11.76 & 11.70 & 11.64 & 11.66 & 12.41 & 8.99 & III \\ 
3045213623404184448 & 07141889-1208085 & 15.35 & 14.20 & 13.62 & 13.51 & 13.29 & 12.72 & 8.99 & III \\ 
3045233998728534272 & 07143643-1159120 & 16.21 & 15.47 & 14.97 & 14.79 & 14.79 & 12.22 & 8.91 & III \\ 
3045233487625692160 & 07135485-1154240 & 16.80 & 15.88 & 15.64 & 15.64 & 15.76 & 11.70 & 8.68 & III \\ 
3045225202636012800 & 07135247-1208309 & 16.23 & 15.27 & 14.69 & 14.68 & 14.51 & 12.05 & 8.92 & III \\ 
\hline
\end{tabular}
\label{tab:IR}
\end{center}
\vspace{-0.3cm}
\end{table*}

\begin{acknowledgements}
We acknowledge support from the Brazilian agencies: 
FAPESP grants 2014/18100-4 (JGH), 2017/19458-8 (AH), and 2014/22095-6 (EM);
and CNPq grants 305590/2014-6 (JGH), and 150465/2019-0 (EM).\\
BL and MdS acknowledge funding from the European Research Council (ERC) under the European Union's Horizon2020 Research and Innovation program for the Project The Dawn of Organic Chemistry (DOC), grant agreement No 741002. 
Based on observations carried out under project number 043-18, 120-18, 034-20 with the IRAM-30m
telescope. IRAM is supported by INSU/CNRS (France), MPG (Germany) and IGN (Spain).
This work has made use of data from the European Space Agency (ESA) mission
{\it Gaia} (\url{https://www.cosmos.esa.int/gaia}), processed by the {\it Gaia}
Data Processing and Analysis Consortium (DPAC,
\url{https://www.cosmos.esa.int/web/gaia/dpac/consortium}). Funding for the DPAC
has been provided by national institutions, in particular the institutions
participating in the {\it Gaia} Multilateral Agreement.
his research has made use of ``Aladin sky atlas" developed at CDS, Strasbourg Observatory, France \citep{ASPC14,AAS00}.
This publication makes use of data products from the Wide-field Infrared Survey Explorer, which is a joint project of the University of California, Los Angeles, and the Jet Propulsion Laboratory/California Institute of Technology, funded by the National Aeronautics and Space Administration. 

\end{acknowledgements}

%
%

\end{document}